# On the Temporal-spatial Analysis of Estimating Urban Traffic Patterns Via GPS Trace Data of Car-hailing Vehicles

Jiannan Mao [a, b] Lan Liu [c], Hao Huang [c], Weike Lu [d, e], Kaiyu Yang [c], Tianli Tang [f], and Haotian Shi [b]*

[a] School of Transportation Engineering, East China Jiaotong University, Nanchang, China;
[b] Department of Civil and Environmental Engineering, University of Wisconsin-Madison, Madison, United States of America;
[c] School of Transportation and Logistics, Southwest Jiaotong University, Chengdu, China;
[d] School of Rail Transportation, Soochow University, Soochow, China;
[e] Alabama Transportation Institute, Tuscaloosa, United States of America;
[f] School of Transportation, Southeast University, Nanjing, China;

*Corresponding author: Haotian Shi, Department of Civil and Environmental Engineering, University of Wisconsin-Madison, 1217 Engineering Hall 1415 Engineering Drive, Madison, WI, United States of America, 53705, e-mail: hshi84@wisc.edu

**Abstract**

Car-hailing services have become a prominent data source for urban traffic studies. Extracting useful information from car-hailing trace data is essential for traffic management, while discrepancies between car-hailing vehicles and urban traffic should be considered. This paper proposes a generic framework for estimating and analyzing urban traffic patterns using car-hailing trace data. The framework consists of three layers: the data layer, the interactive software layer, and the processing method layer. By pre-processing car-hailing GPS trace data through operations such as data cutting, map matching, and trace correction, the framework generates tensor matrices that estimate traffic patterns for car-hailing vehicle flow and average road speed. An analysis block, built on these matrices, examines the relationships and differences between car-hailing vehicles and urban traffic patterns, aspects that have been overlooked previously. Experimental results demonstrate the effectiveness of the proposed framework in examining temporal-spatial patterns of car-hailing vehicles and urban traffic. For temporal analysis, urban road traffic displays a bimodal characteristic, whereas car-hailing flow exhibits a 'multi-peak' pattern, fluctuating significantly during holidays and thus creating a hierarchical structure. For spatial analysis, the heat maps generated from the matrices exhibit discrepancies, but the spatial distribution of hotspots and vehicle aggregation areas remains similar.

**Keywords**: Intelligent Transportation Systems; Car-hailing; GPS trace data; Temporal and spatial distribution;

# 1. Introduction

Urban traffic patterns embody the intricate and dynamic interplay between vehicular movements and human activities within road networks. Understanding and accurately estimating these patterns can substantially enhance traffic efficiency, safety, and sustainability in urban areas(Correa & Ozbay, 2022). Recent advancements in communication technology and fundamental infrastructure have facilitated the acquisition of Global Positioning System (GPS) data from Intelligent Traffic Systems (ITS) (Lu et al., 2023; J. Yu et al., 2020). These supply traffic management and control systems with a more comprehensive understanding of vehicle trajectories compared to conventional sensor data (D'Andrea & Marcelloni, 2017; He et al., 2019). In particular, GPS trace data can be employed to estimate urban traffic patterns effectively (Herrera et al., 2010). Consequently, extracting valuable spatial-temporal information from GPS trace data and analyzing traffic patterns derived from such data has become a topic of considerable academic importance.

GPS data can be collected via vehicle onboard units or mobile phone data to illustrate vehicle movements (Tu et al., 2021). Mining the GPS data can extract traffic metrics (e.g., traffic speed and flow) and then analyze pertinent information, such as temporal distribution patterns and spatial aggregation states of traffic (H. Liu et al., 2024; Lu et al., 2023; Jiechao Zhang et al., 2022). In this domain, prior research has employed GPS trace data for various purposes. For example, Moreira et al. (Moreira-Matias et al., 2013) predicted the spatial distribution of taxi passengers using taxi GPS traces to devise a more intelligent taxi dispatching strategy. An et al. (An et al., 2018) identified recurring congestion evolution patterns by employing grid divisions and GPS-equipped vehicle mobility data. Several studies have also integrated trace data with traffic flow theories to examine traffic patterns. For instance, using GPS data, Yu et al. (Wang Yu et al., 2022) developed a comprehensive model incorporating a real-time lane identification model and a real-time queue length estimation model based on traffic shockwave theory. Similarly, Tu et al. (Tu et al., 2021) combined three-phase traffic flow theory and presented a framework to categorize traffic flow states, which can process massive, high-density, and noise-contaminated smartphone sensor data sets.

Although the wealth of trace data has profoundly influenced numerous studies in the traffic domain, significant challenges persist in utilizing vehicle trajectory data for traffic estimation tasks. Specifically, original GPS traces require additional processes such as map-matching for traffic estimations and also necessitate an efficient framework for accurate large-scale urban traffic speed calculation (Herrera et al., 2010; Tu et al., 2021; J. Yu et al., 2020; Wenhao Yu, 2018). Moreover, some coarse-grained algorithms, primarily designed for business applications, may not be suitable for road-based estimation tasks since they partition urban networks into grids and allocate different traces to each grid (An et al., 2018; Liu et al., 2020; Wenhao Yu, 2018). Consequently, it is crucial to develop a practical road-level traffic pattern estimation framework capable of addressing incomplete data, such as data without instantaneous speed measurements.

Additionally, recent studies have focused on developing specialized approaches for distinct technical domains using GPS data, such as map-matching (D'Andrea & Marcelloni, 2017)(Hu et al., 2023), semantic information interpretation (Al-Dohuki et al., 2016; Liu et al., 2020), and incomplete matrix completion (X. Chen et al., 2019; J. Yu et al., 2020; Jinlei Zhang et al., 2024). Further, some studies dedicated efforts towards the analysis of network traffic patterns through GPS data, including macroscopic traffic patterns (Y. Zhang et al., 2023), visualization of congestion causes (Pi et al., 2019), and the spatial-temporal distribution of congestion (H. Liu et al., 2024; Jiechao Zhang et al., 2022). While these studies have demonstrated efficacy in addressing specific needs and requirements, the increasingly rich source of car-hailing data necessitates a concise and effective processing framework. Car-hailing fleets leverage smartphone sensor data to avoid complex installations and high maintenance costs, providing extensive spatial and temporal coverage that results in high user penetration rates (J. Liu et al., 2024). However, despite the extensive use of car-hailing data, there remains a gap in the literature concerning the comparative analysis of car-hailing services and the urban traffic patterns estimated from this data, which is crucial for a comprehensive understanding of urban mobility dynamics.

In addition to addressing traffic estimation concerns (Correa & Ozbay, 2022; Jinlei Zhang et al., 2024), a generic framework is also desirable to connect potential applications in academia and car-hailing service providers (e.g., Uber, Lift, and DiDi). How can academic researchers conveniently utilize the data provided by car-hailing service companies to generate urban traffic estimations? And how can car-hailing service providers benefit from the academic outputs cost-effectively? It is essential to consider the trade-off between business usage and academic research, and accommodate the needs of both parties.

## 2. Research goal and objectives

With the above considerations, this paper proposes a generic traffic pattern estimation and analysis framework utilizing car-hailing vehicle traces. The proposed estimation framework comprises three layers: the data layer, the software interaction layer, and the processing method layer. The data layer is designed to accommodate car-hailing trajectory data, while the software interaction layer integrates Python[1], Open Streat Map (OSM)[2], and ArcGIS[3] to jointly process trace data and network information data. The processing method layer employs two types of data processing techniques: map correcting techniques, including map-matching and trace corrective action, and data pre-processing techniques, such as data cutting and traffic measurement calculation. This estimation framework is designed for easy implementation with minimal software and data source requirements and can be further adapted to suit the evolving needs of various applications.

Next, we evaluate the performance of the proposed estimation-analysis framework using real-world car-hailing data. With car-hailing trace data as input, the car-hailing vehicle flow and the mean speed of road segments are calculated every fifteen minutes. The INRIX index (Dong et al., 2019; Reed, 2019), a widely used measure of road congestion levels, is chosen to assess the congestion level of roads. On this basis, we conduct comprehensive analyses of the temporal-spatial patterns of car-hailing vehicles and urban traffic, elucidating the relationships and differences between them.

In summary, the main contributions of this paper are in two aspects. (i) We introduce a generic framework for traffic estimation using car-hailing trace data, adaptable to various practical situations. The estimation framework calculates car-hailing flow and urban traffic speed matrices at the road level, which are instrumental for advancing both fine-grained transportation management and academic studies. (ii) The analysis block of our framework reveals relationships between car-hailing vehicles and urban traffic patterns in the temporal-spatial domain, which are valuable for enhancing car-hailing service management and broader applications. Additionally, we address the often overlooked disparities between car-hailing vehicles and urban traffic patterns, a crucial factor for accurate urban traffic assessment when relying on car-hailing service data.

The rest of this paper is organized as follows: Section 3 introduces the specific content of the estimation-analysis framework and evaluation metrics; Section 4 shows real-world experimental results based on car-hailing data. Section 5 illustrates temporal and spatial distribution patterns derived from traffic speed and car-hailing vehicle flow, respectively; Section 6 discusses the difference between car-hailing vehicles and urban traffic flow; Section 7 concludes the findings

---

[1] https://www.python.org
[2] https://www.openstreetmap.org/
[3] https://developers.arcgis.com

and introduces future works.

## 3. Methodology

### *3.1 Technical architecture*

Based on the objectives discussed in Section 2, this study aims to: (i) estimate car-hailing vehicle flow and urban road speed simultaneously using GPS data of car-hailing vehicles, and (ii) analyze temporal-spatial patterns of car-hailing service and urban traffic, particularly disparities in these two patterns.

To achieve these objectives, we developed a framework for estimating and analyzing car-hailing fleets and urban traffic, as illustrated in Figure 1. Specifically, the estimation framework comprises three primary layers: the data layer for standardizing the input data, the interactive software layer for establishing an underlying software foundation, and the processing method layer for handling pre-processing and estimating tasks. An analysis block is built upon the estimation results to examine the temporal-spatial patterns of car-hailing services and urban traffic vehicles. It is important to note that the processing method layer serves as a functional layer that closely interacts with the other two layers. Details of each layer will be discussed in the subsequent sub-sections, and the primary code can refer to https://github.com/gxwzysjdkevin/gps_user_demo.git.

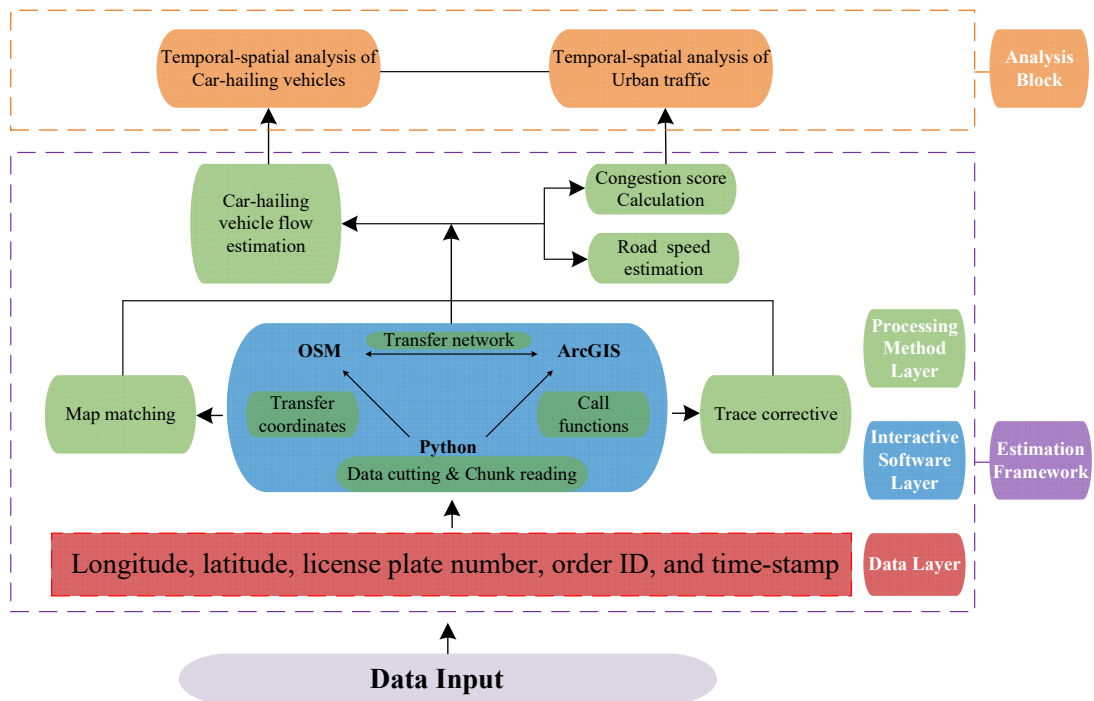

**Figure 1.** Flow chart of the estimation framework and analysis block

## 3.2 Data layer and interactive software layer

The data structure in the data layer should contain the longitude, latitude, anonymized driver and order ID, and time stamp of each point. For large-scale urban networks, the daily data magnitude will be in gigabytes (GB), which requires an efficient data reading and processing technique. Regarding the software interaction layer, we utilize Python, OSM, and ArcGIS to handle data processing and network structure representation.

These two layers aim to build a data and software environment for subsequent processing. For efficient data reading, chunk reading with a size of 10,000 is chosen to add data to files. Chunk reading can handle car-hailing data at an urban scale, as most car-hailing trace data, in reality, is in the GB magnitude. Regarding road map illustration and map matching of trace data, OSM can provide abundant urban network data, serving as a map base in further processing.

## 3.3 Processing method layer

The processing method layer is an essential part of the traffic pattern estimation framework, responsible for the necessary processing techniques and methods required for pre-processing data and map-matching operations. Specifically, its functions include data cutting, ArcGIS operation, car-hailing vehicle flow estimation, road average speed estimation, and congestion score calculation.

### 3.3.1 Data cutting

The first function of the processing method layer is data cutting, which involves transferring the read data to facilitate loading. As mentioned in Section 3.2, data is loaded using a chunk reading size of 10,000, and a sampling interval of fifteen minutes is set to divide the read data into sample pieces. The purpose of this function is to minimize memory usage and provide a database for illustration and analysis. In this study, we found that a sampling interval of fifteen minutes effectively captures the overall evolution of traffic in urban road networks for broader analysis. However, it is important to note that this interval may not be suitable for all applications. For instance, real-time routing or precise estimation of travel times would likely require much shorter sampling intervals to accurately reflect dynamic traffic conditions.

### 3.3.2 Map operation

The map operation is employed for trace correction and map-matching, using a two-step approach to enhance accuracy. This offline operation initially corrects the coordinate shift by estimating local deviations and then maps the trajectory points to roads with the shortest distances and labels the points for the corresponding road. The map operation can be handled by the 'Near_Analysis' function in ArcMap, a combined tool of ArcGIS and Python, for fine-tuning. This intuitive and user-friendly function conducts near-neighbor analyses that refine the trace alignment by matching trajectory points to the closest roads on the OSM base map, based on the shortest path criteria.

As the sampling rates and data quality improve, the accuracy of this point-to-curve operation, essential for ensuring precise map-matching, reaches an acceptable range (Quddus & Washington, 2015). The enhanced map operation not only allows users to effectively tackle trace-correction and map-matching challenges but also provides an interactive interface for engineering-focused applications. To achieve potentially higher accuracy and overcome measurement noise, more advanced techniques can be applied, as suggested by recent studies (C. Chen et al., 2019; Dogramadzi & Khan, 2021; D. Zhang et al., 2021).

*3.3.3 Car-hailing vehicle flow and road speed estimation*

One aspect of our estimation process focuses on determining the flow of car-hailing vehicles. We achieve this by utilizing a fifteen-minute sample interval calculation, which results in a tensor matrix containing the estimated number of car-hailing vehicles per road segment for each sample interval. Generally, each car-hailing order rarely passes the same road section twice within a short period due to factors such as fuel cost and profit optimization. Consequently, the car-hailing vehicle flow estimation is approximated by counting the order IDs of different car-hailing vehicles that accumulate on each road segment during each sample interval.

Another aspect of the estimation process involves estimating the average speed of road sections using car-hailing vehicles as indirect indicators. The underlying principle is that the average speed of a road section can be indirectly inferred from car-hailing vehicles that travel through it during a specified sample period. As a result, the calculation procedure can be structured as follows. First, pick out different traces according to order IDs and sort the points in each trace in terms of time stamps; then, the average speed can be viewed as the ratio of the nearest distance (calculated by latitude and longitude of trajectory points) to the difference of time-stamp. Detailed calculation steps are listed below:

First, the Haversine formula (Gade, 2010) is employed to compute the spherical distance $d$ between any pair of longitude and latitude points. The formula is as follows:

$$d_{r_{ij} \subset \mathbf{r}_k} = 2r \arcsin\left(\sqrt{\sin^2(\frac{\varphi_j^{lat} - \varphi_i^{lat}}{2}) + \cos\varphi_i^{lat} \cdot \cos\varphi_j^{lat} \cdot \sin^2(\frac{\lambda_j^{lon} - \lambda_i^{lon}}{2})}\right) \quad (1)$$

where, $d_{r_{ij} \subset \mathbf{r}_k}$ is the distance of the nearest trace pair $r_{ij}$, i.e., the distance between two nearest points $i$ and $j$ of the same GPS trace; $\mathbf{r}_k$ denotes the nearest GPS trace pair set that is located on road $k$, and $r_{ij} \subset \mathbf{r}_k$; $r$ is the radius of the sphere(in this work, the approximate radius of the earth, 6378.137 km); $\varphi_i^{lat}$ and $\varphi_j^{lat}$ are the latitude of GPS trace points $i$ and $j$, respectively; Analogous, $\lambda_i^{lon}$ and $\lambda_j^{lon}$ are the longitude of GPS trace points $i$ and $j$, respectively.

Then the speed of the trace point $r_{ij}$ during the sample period $\Delta t_{ij}$ (the sample period is chosen according to time stamps of the nearest two points with a threshold of 10 seconds) can be computed as follows:

$$v_{r_{ij} \subset \mathbf{r}_k}^{\Delta t_{ij}} = \frac{d_{r_{ij} \subset \mathbf{r}_k}}{\Delta t_{ij}} \quad (2)$$

where $v_{r_{ij} \subset \mathbf{r}_k}^{\Delta t_{ij}}$ is the speed of the trace pair $r_{ij}$.

Note that the calculation in Eq. (1) and (2) may cause loss of accuracy when dealing with lower sampling frequencies of GPS data, as straight-line distances between infrequent points may not accurately represent the actual road path. In this circumstance, additional features such as heading directions and ensemble with shortest path algorithms (C. Chen et al., 2019; Hsueh & Chen, 2018; Quddus & Washington, 2015) can help to modify this misestimation. Here, we only provide a simplified version, as shown in Eq. (1) and (2), and it is efficient in high sampling frequency data.

Based on Eq.(1) and Eq. (2), the average speed of road $k$ can be obtained by computing the arithmetic mean of velocities of trace GPS trace pair set $\mathbf{r}_k$ as follows:

$$\bar{v}_{\mathbf{r}_k} = \frac{1}{n(k)}\sum_1^{n(k)} v_{r_{ij} \subset \mathbf{r}_k}^{\Delta t_{ij}} = \frac{1}{n(k)}\sum_1^{n(k)} \frac{d_{r_{ij} \subset \mathbf{r}_k}}{\Delta t_{ij}} \quad (3)$$

where $\bar{v}_{\mathbf{r}_k}$ is the average speed of road $k$; $n(k)$ is the total number of nearest GPS trace pairs that belong to road $k$; Note that the road speed is set to 0 when there is no GPS trace during the sample period.

Also note that an anomaly threshold will be set to test if the estimated speed is wrong due to map-matching errors such as sudden trace shifts. The anomaly threshold can be determined as the speed limit in urban regions. If the speed exceeds the threshold value, the speed can be substituted by the interpolation of its neighborhood periods.

This framework generally provides a straightforward yet effective method for estimating the car-hailing vehicle flow and road speed. While more sophisticated estimation methods may exist, our framework focuses on achieving computational efficiency and relatively high accuracy. As such, this framework serves as a practical solution for engineering-driven tasks.

*3.4 Analysis block*

The analysis block conducts temporal-spatial analyses for both car-hailing vehicles and urban traffic. It investigates their temporal and spatial dynamics(e.g., temporal patterns and spatial distribution) and compares their characteristics and potential implications.

Modifying the estimated traffic speed matrix is necessary for the temporal-spatial analysis of urban traffic, given the variability of traffic speeds across different roads. One straightforward way to present the traffic state is to evaluate traffic congestion on the same scale. In this work, to simplify the evaluation process, we measure the congestion level solely based on the average road speed estimated by GPS traces. Therefore, we adopt INRIX index(Reed, 2019), which assesses the congestion degree of roads by average road speed, where a lower score indicates a better condition. The INRIX index measures congestion values by using the actual and free-flow speed of a certain road section, according to the following formula:

$$\text{INRIX}_{kl} = \begin{cases} (\frac{\text{TH}_k}{\text{RE}_{kl}}-1), & (\frac{\text{TH}_k}{\text{RE}_{kl}}-1) > 0 \\ 0, & (\frac{\text{TH}_k}{\text{RE}_{kl}}-1) \leq 0 \end{cases} \quad (4)$$

where, $\text{INRIX}_{kl}$ is the congestion score of the road $k$ during the *l-th* sample interval; $\text{TH}_k$ is the free flow speed of road $k$; $\text{RE}_{kl}$ is the average actual speed of road $k$ during the *l-th* sample interval.

When evaluating network-wide congestion levels, a weighted calculation can be applied as follows:

$$\text{INRIX}_l = \frac{\sum L_k \times \text{INRIX}_{kl}}{\sum L_k} \quad (5)$$

where, $\text{INRIX}_l$ is the network congestion score during the *l-th* sample interval; $L_k$ denotes the length of road $k$.

**4. Experimental settings and processing**

*4.1 Experimental settings and map operation*

This section presents the experiment setting for evaluating the proposed three-layer estimation framework and analysis block, using anonymous car-hailing trace data from a part of Chengdu, China. The study area contains the central CBD of Chengdu city and is about 59.5 km$^2$. The records include 30 days of car-hailing vehicle traces from October 1$^{st}$ to October 30$^{th}$, 2016, with approximately 40 million daily trace records. More specifically, each row of the data contains anonymized driver ID, order ID, timestamp (Unix timestamp in seconds), and the corresponding latitude and longitude (GCJ coordinates).

The car-hailing trajectory points were adjusted to be mapped into OSM coordinates using Map operations. The result of a specific part of the experimental region is shown in Figure 2. The light green lines represent the base map obtained from OSM, while the dark green points are the car-hailing trajectory points. An amplified of the city center is presented for a clear comparison. As shown in Figure 2(a), the original traces and OSM base map deviate but can be fixed by a shift in the northwest direction, as indicated by a yellow arrow. Then basic ArcGIS operations are carried out to rectify trajectory points. As shown in Figure 2(b), it is clear that the operation is sufficient for correcting the shift and mapping the traces to corresponding roads.

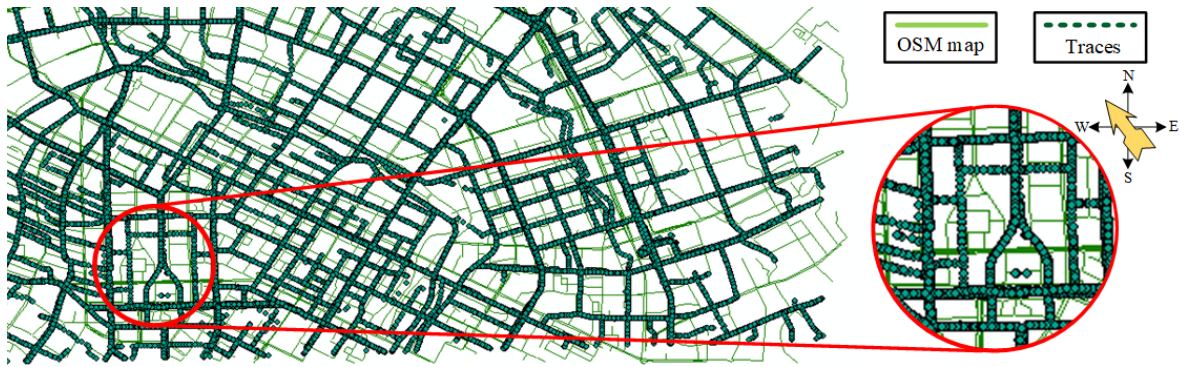

(a) Before correcting. The yellow arrow denotes the correction operation direction.

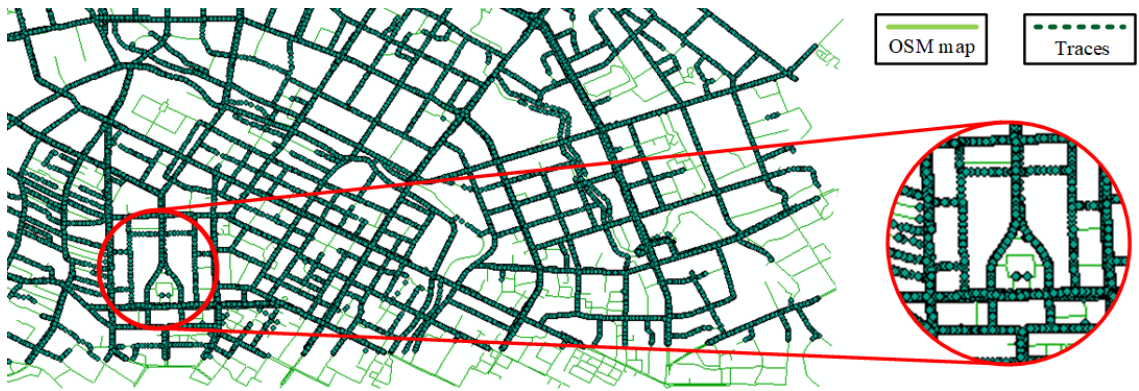

(b) After correcting. Traces overlap well with the base OSM map.

**Figure 2.** Corrective operation with car-hailing trajectory points and the OSM map

*4.2 Estimation process and data clean*

Using the estimation framework, GPS trajectory data can be converted into two tensor matrices. Besides, the mean speed matrix requires further data pre-processing because the road speed data may only be collected when a road order record exists. As the study area is the central part of Chengdu city, it stands to reason that a busy central area in a large modern city will be the hotspot for car-hailing services. Therefore, we believe that removing those roads with a high fraction of missing values will have little effect on analyzing urban traffic patterns.

Here, we set a threshold of 0.2 to filter missing values and about 1,000 roads left. A simple linear interpolation is used to complement the rest of the data with a small number of missing values. Further, the anomaly threshold is set as 70 km/h according to the speed limit of China. Estimated speed values exceeding this threshold will be replaced by the interpolation of its neighborhood periods. These anomalies count for less than 0.1% of total values in the spatial-temporal matrix.

*4.3 Degree of congestion (DC) and car-hailing flow (CF)*

As described in the methodology section, we employ the INRIX scores derived from the speed tensor matrix to measure the degree of traffic congestion (DC), which is a dimensionless measurement in the range of [0,1]. The car-hailing service is measured by car-hailing vehicle flow (CF), denoting the number of car-hailing vehicles per sample period. For a clear comparison version, the network-scale mean value of DC and the total value of CF on roads are computed as network-wide measurements. Then, we employ a fitting index $f^2$ to quantify the dispersion degree of data, which can be computed as follows:

$$f^2 = 1 - \frac{\sum_{j=1}^{n}(y_j - \bar{y}_j)^2}{\sum_{j=1}^{n}(y_j - \bar{Y})^2} \tag{6}$$

where, $y_j$ is the actual network-wide value (flow/speed) of sample interval $j$; $\bar{y}_j$ is the mean of all actual values during sample interval $j$; $\bar{Y}$ denotes the mean value of all $n$ sample intervals; The closer $f^2$ is to 1, the greater the number of values aggregated together, indicating that traffic state patterns are similar each day.

It should be noted that the arithmetic mean of instantaneous GPS speed can reflect the whole traffic condition of the road section, while the car-hailing vehicle flow only stands for car-hailing vehicle service. Thus, it is possible to mine the relationship between car-hailing vehicles and urban traffic patterns by comparing the two measurements, DC & CF.

## 5. Temporal-spatial pattern analysis

The following subsections will conduct temporal and spatial pattern analyses based on the CF of car-hailing services and the DC of road traffic. By examining these patterns, we aim to uncover the inherent correlations between car-hailing services and urban road traffic, which can provide valuable insights into urban traffic dynamics.

### *5.1 Temporal pattern analysis*

This section analyzes temporal patterns of car-hailing vehicles and road section traffic, and comparisons will be made to excavate the relationship between them. The DC and CF for each sample interval are plotted in Figure 3. The chosen data period includes a seven-day holiday (from October 1$^{st}$ to October 7$^{th}$). In Figure 3, each blue point represents the network-wide value every fifteen minutes. The x-axis represents the time of day, while the y-axis denotes the DC/CF values. The sub-plots, Figure 3(a-c), illustrate the DC, whereas sub-plots, Figure 3(d-f), depict the CF.

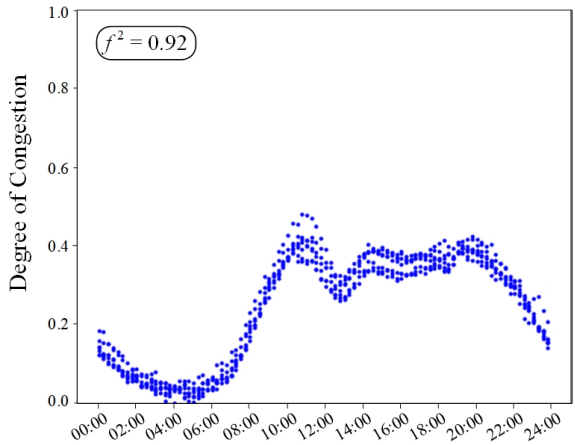
(a) National Day (DC)

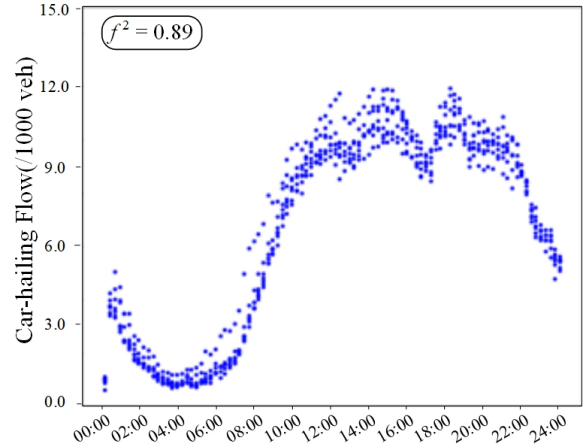
(d) National Day (CF)

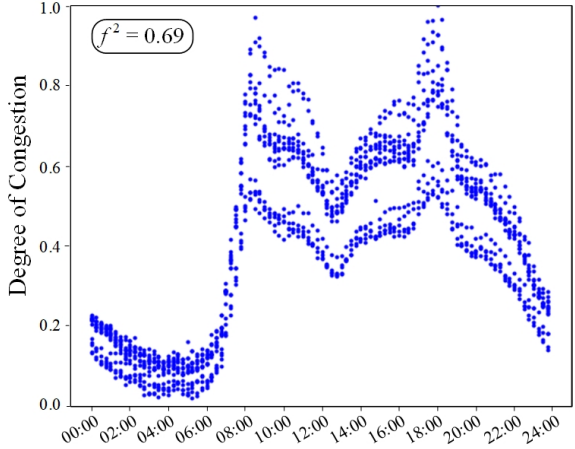
(b) Weekday (DC)

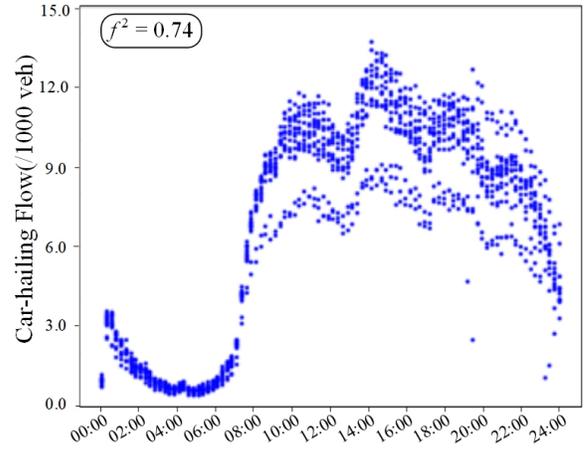
(e) Weekday (CF)

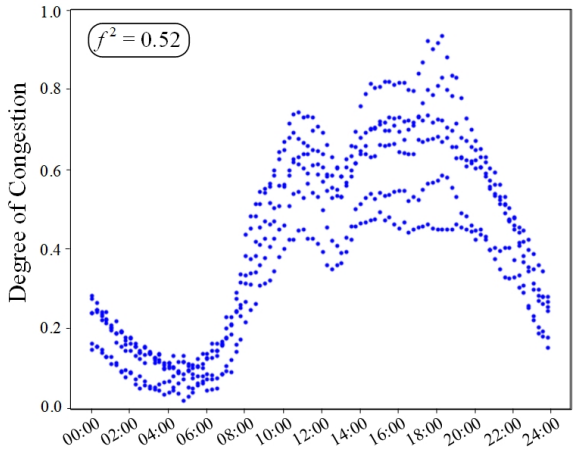
(c) Weekend (DC)

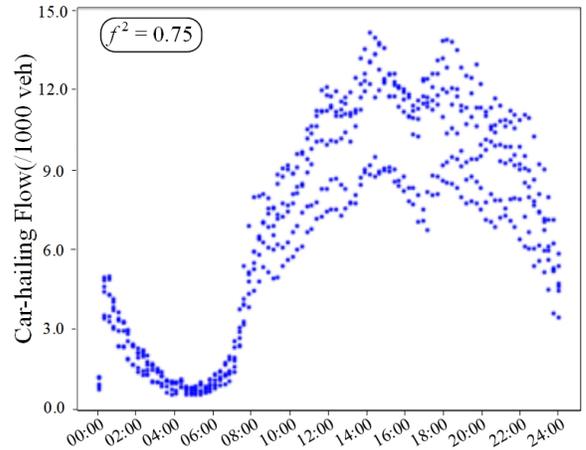
(f) Weekend (CF)

**Figure 3.** Degree of congestion (DC) and car-hailing Flow (CF) for different periods

In Figure 3(a-c), a bimodal distribution can be observed, and the congestion degree is low on the holiday (Figure 3(a)). The low level of congestion can be attributed to the following: (i) a large number of urban residents travel on vacation to other cities, relieving the daily traffic pressure in the central part of Chengdu; (ii) commuting peaks decrease during holidays and the travel time distribution on the network becomes homogeneous, thereby reducing the level of congestion. As for car-hailing flow in Figure 3(d-f), a 'multi-peaks' characteristic appeals on holidays, weekdays, and weekends. Traffic states of these multi-peaks will switch on and off between the congestion and steady flow state (Kerner et al., 2007). And this 'multi-peaks' characteristic, unaffected by holidays or weekends, is triggered by increased demand for car-hailing services.

Notably, an interesting hierarchical structure appears in both DC and CF plots, except for holidays. The fitting index $f^2$ also reveals this phenomenon mathematically. $f^2$ of the two measurements are larger than 0.9 during holidays, while during other times, they range between 0.52 and 0.75. The hierarchical structure suggests that daily travel conditions fluctuate, whether for car-hailing or urban roads. Figure 4 depicts the total CF and average daily DC of the network, where the x-axis represents the date, and the y-axis represents the measurement value. The red and blue lines indicate measurements of weekdays and weekends or holidays, respectively.

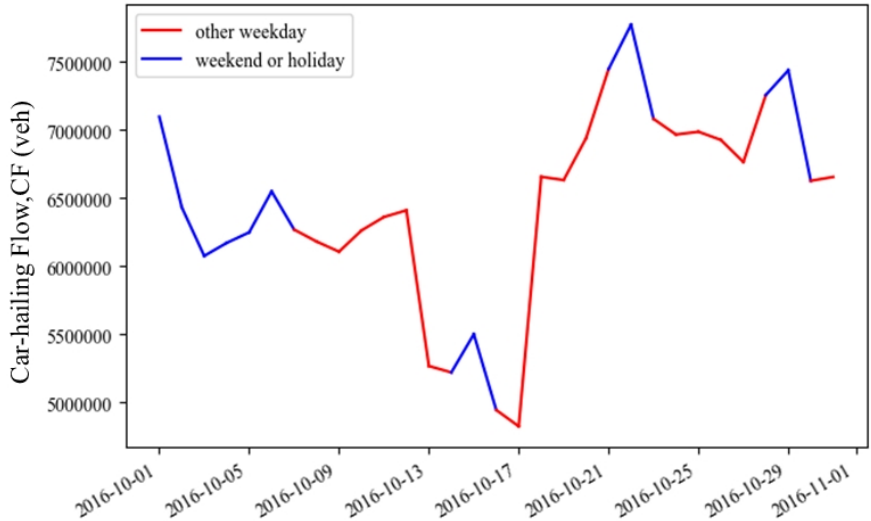

(a) Total car-hailing flow per day

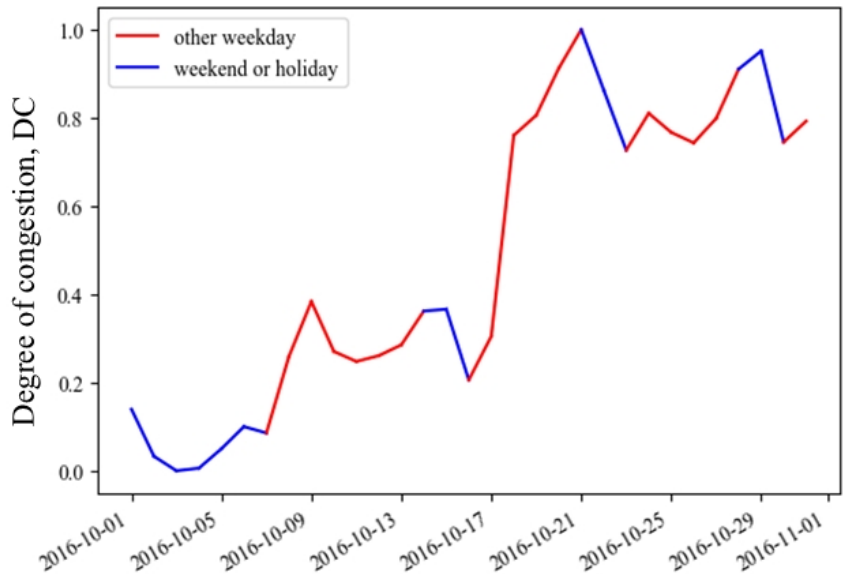

(b) Degree of congestion per day

**Figure 4.** Total daily CF and mean DC of the network

It can be seen from Figure 4 that a flow drop (about 12$^{th}$ to 18$^{th}$ October) and a congestion degree rise (about 16$^{th}$ to 18$^{th}$ October) after the National Day holiday. As there were no extreme weather conditions and big public activities during this period, this may be the reason for the appearance of hierarchical structures. Following this step, a min-max normalization calculation is performed to eliminate the impact of fluctuation as follows:

$$x_N = \frac{x_j - \min(\mathbf{x}_{day})}{\max(\mathbf{x}_{day}) - \min(\mathbf{x}_{day})} \tag{7}$$

where, $x_N$ is the min-max normalization value of $x_j$ of sample interval $j$; $x_j$ is the original measurement value at sample interval $j$; $\mathbf{x}_{day}$ denote measurement value set of all sample intervals in a certain day.

The min-max normalization values of DC and CF, i.e., the DC-N and CF-N values, are illustrated in Figure 5. In all scenarios, the hierarchical structures disappear, and the fitting index values reach quite a high level, at around 0.89~0.95. Consequently, we can deduce that the fluctuation of measurement values causes hierarchical structures. However, a crucial point still requires further explanation: what causes this fluctuation?

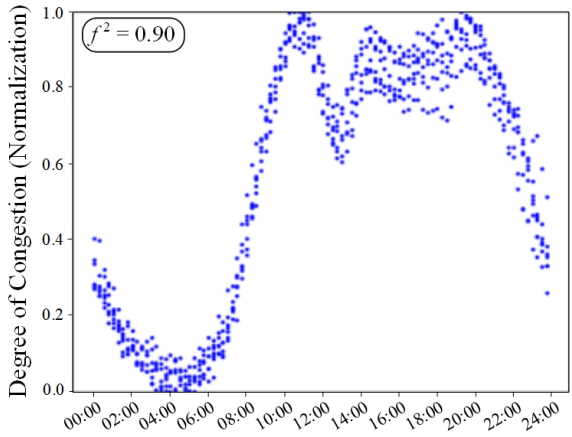

(a) National Day (DC-N)

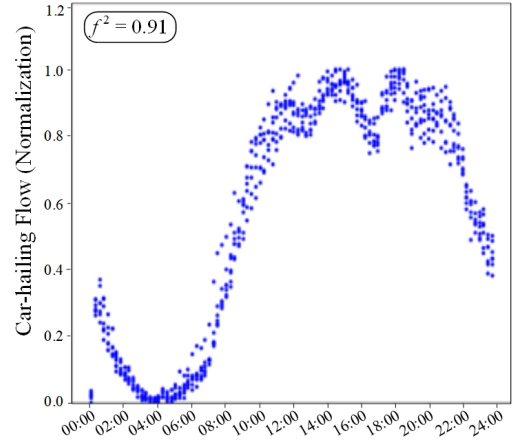

(d) National Day (CF-N)

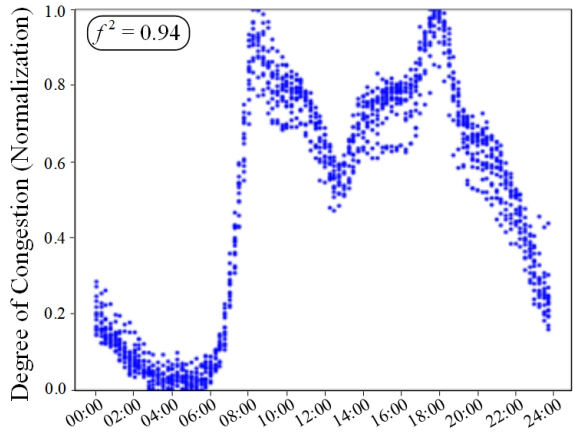
(b) Weekday (DC-N)

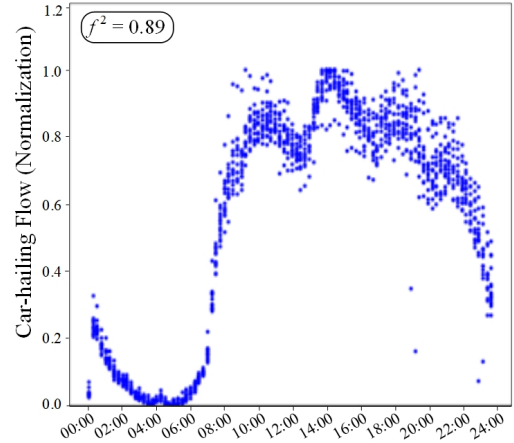
(e) Weekday (CF-N)

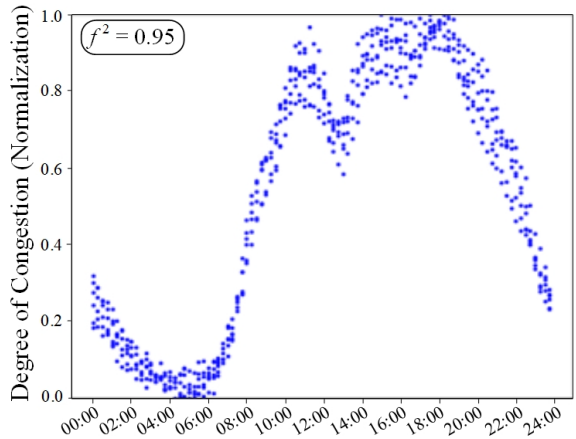
(c) Weekend (DC-N)

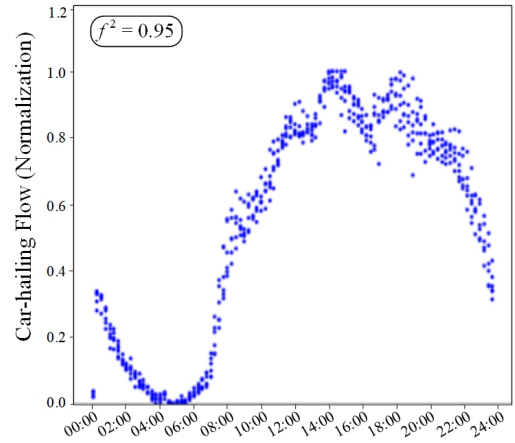
(f) Weekend (CF-N)

**Figure 5.** Daily car-hailing flow and INRIX score values (after normalization)

In terms of the general trend shown in Figure 4, the degree of congestion remains low during holidays, while the car-hailing flow is at a rather high level. It demonstrates that car-hailing accounts for a bigger share of travel modes during holidays than usual, consistent with holiday travel behaviors. Therefore, the number of car-hailing vehicles decreases after the holidays, as car owners tend to accept car-hailing orders during the holidays and then rest **after the holidays** to maximize revenues. This number decline will lead to fluctuation and cause hierarchy. Regarding road traffic, the main reason is that companies shift vacations a few days before or back. Different companies will have various vacation shift schedules, resulting in a delayed increase in congestion degree. After approximately one week, daily traveling returns to normal, and traffic volume rises, generating fluctuations.

In conclusion, the car-hailing vehicle flow displays a 'multi-peaks' characteristic, resulting from service demands across multiple periods. This characteristic is somewhat diminished during holidays, suggesting that car-hailing vehicles are directly influenced by the alternation between workdays and holidays. Inferred from the estimated urban traffic speed data, the degree of congestion values exhibit a distinct bimodal characteristic. This is consistent with traffic flow features, thereby validating the effectiveness of the proposed estimation framework.

*5.2 Spatial pattern analysis*

The formation and dissipation of car-hailing vehicle flow and urban road traffic congestion scores are observed hourly throughout National Day, weekdays, and weekends. Several critical transitional points are selected and displayed in Figure 6. The label denotes the ratio to the maximal measurement values. Figure 6(a-d) depict the spatial distribution of car-hailing vehicles from 2-5 a.m., 7-8 a.m., 11-12 a.m., and 7-8 p.m., whereas Figure 6(e-h) depict the spatial distribution of urban road traffic of 2-5 a.m., 7-8 a.m., 9-10 a.m., and 9-10 p.m. It should be noted that a major difference between the two types of figures is that the heatmap of CF volumes is much sparser than that of DC, which is caused by the magnitude of measurement methods. The difference will not affect the main pattern of each type of heatmap.

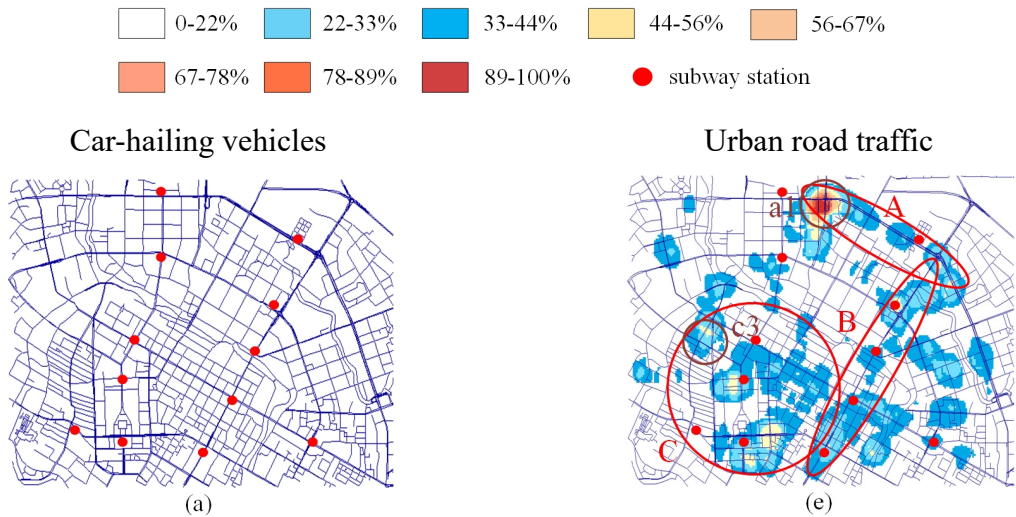
18

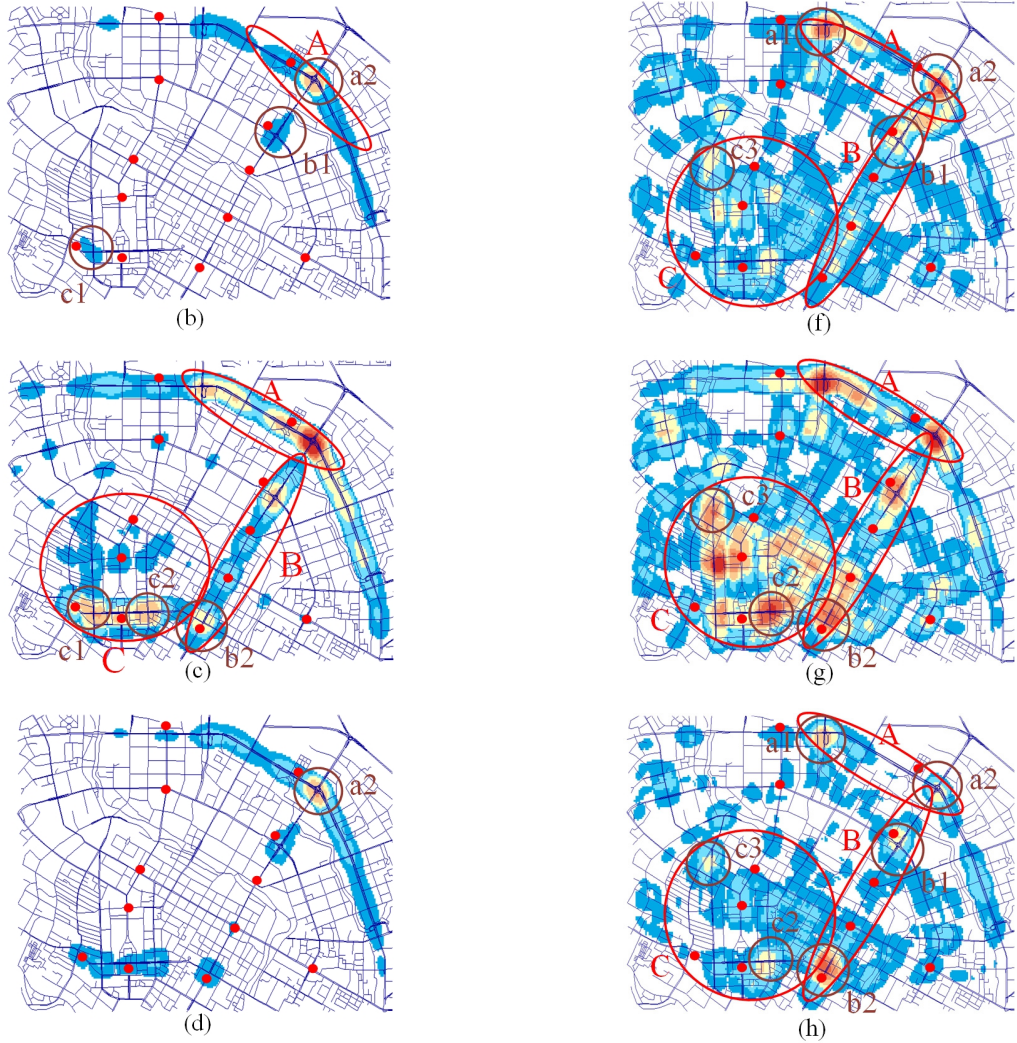

**Figure 6.** Spatial distribution of car-hailing vehicles and urban road traffic.

As in Figure 6(a), for CF volume, all roads are in low values. However, for the urban road traffic in Figure 6(e), the *a1* intersection in area *A* has a high congestion score. Several roads in areas *A*, *B*, and *C* have relatively high congestion scores, with vehicles primarily concentrated around subway stations, which can be considered commercial centers.

The volume of CF increases gradually after 6 a.m., as demonstrated in Figure 6(b). Car-hailing vehicles congregate on the ring road of area *A*, with a high value at point *a2* and a moderate clustering at *b1*, *c1*. For DC, the congestion in area *A* dissipates outward from *a1* and *a2* at 7 a.m., as depicted in Figure 6(f). Analogous, multiple subway stations located in areas *B* and *C* have relatively high congestion scores, which radiate from subway stations.

Since then, the number of car-hailing vehicles continued to increase, as shown in Figure 6(c).

The high-value area of the ring road in area *A* continues to expand around *a2*, while new high-value clustering *b2* and *c2* appear in regions of *B* and *C*. Additionally, numerous intersections begin to exhibit clustering phenomena. This situation persists till 11 a.m. and is virtually unchanged from 12 a.m. to 7 p.m. After 7 p.m., the aggregation of car-hailing vehicles begins to dissipate slowly. As shown in Figure 6(d), the aggregation areas *A*, *B*, and *C* shrink to *a2*, *b1* and *b2*, *c2*, respectively. The status will return to Figure 6(a) at around 2 a.m.

On the other hand, as urban road traffic congestion levels reach the peaks at 8 a.m., DC at *b2* and *c2* also grow significantly, as illustrated in Figure 6(g). Then the spatial distribution pattern can be divided into two categories: (i) from 9 a.m. to 6 p.m on the holidays and weekends, the spatial distribution of DC is relatively stable; (ii) from 9 a.m. to 1 p.m. on weekdays, congestion gradually dissipates and forms the spatial distribution pattern depicted in Figure 6(f). DC values continue to increase till 6 p.m., and the traffic congestions in areas *A*, *B* and *C* are still centered on neighbors of *a2*, *b1* and *b2*, *c1* and *c2*. Then, road congestion decreases, and at around 9 p.m., the spatial distribution pattern turns to the status shown in Figure 6(h). It will then revert to Figure 6(e) for the next day.

The aggregation areas of car-hailing vehicles are mainly located in three areas: the ring road of area *A*, the main streets of area *B*, and the large square area of the city center in area *C*. The aggregation points are *a2*, *b1*, and *c1* before 11 a.m., while after 11 a.m. *b2* and *c2* emerge. Car-hailing vehicles tend to congregate around intersections and are impacted by the distribution of subway stations. As for urban traffic congestion, easy-to-congest areas are also in areas *A*, *B*, and *C*. The aggregation point is located at *a1* before 5 a.m. During 5 and 9 a.m., vehicles congregate at subway stations, and two new aggregation points, *b2* and *c2*, appear after 9 a.m. In addition, there is a modest aggregation at *c3*, although no significant congestion occurs.

In summary, while there are similarities between CF and DC heat maps, with shared hotspots in areas *A*, *B*, and *C*, some differences also exist. Furthermore, the vehicle aggregation points for the two heatmaps exhibit a certain correlation: gathering around *a2*, *b1*, and *c1* before morning peaks, and then generating two new points (*b2* and *c2*).

## 6. Discussions on differences between car-hailing vehicles and urban traffic flow

### 6.1 Temporal differences

Reliable assessments of urban traffic speed can be made by GPS-enabled car-hailing services (Herrera et al., 2010). However, it does not imply that car-hailing vehicles and urban traffic patterns will be similar—they exhibit different daily forms. Differences can be seen from the result of temporal pattern analysis, as shown in Figure 3 and Figure 5. A general sample diagram of the two patterns is illustrated in Figure 7 for easy comparison.

In Figure 7, the y-axis is the dimensionless measurement values of CF and DC, while the x-axis is the time of the day. Note that the daily sample diagram is drawn according to the general shapes of DC and CF in Figure 3 and Figure 5. For car-hailing vehicles (CF, shown as orange dot-dash lines in Figure 7), additional peaks occur despite the two common peaks (morning and evening peaks, denoted as red arrows) as in urban traffic (DC, shown as blue lines in Figure 7). One is at around 2 p.m., and the other is at around 9 p.m., denoted as orange arrows. We regard these peaks as products of different travel behaviors of car-hailing vehicles: unlike common daily commute purposes, customers tend to employ car-hailing services in the early afternoon for entertainment purposes and at around 9 p.m. for returning home. The additional peaks in car-hailing services are the main points of difference compared to urban vehicle patterns in the temporal scope.

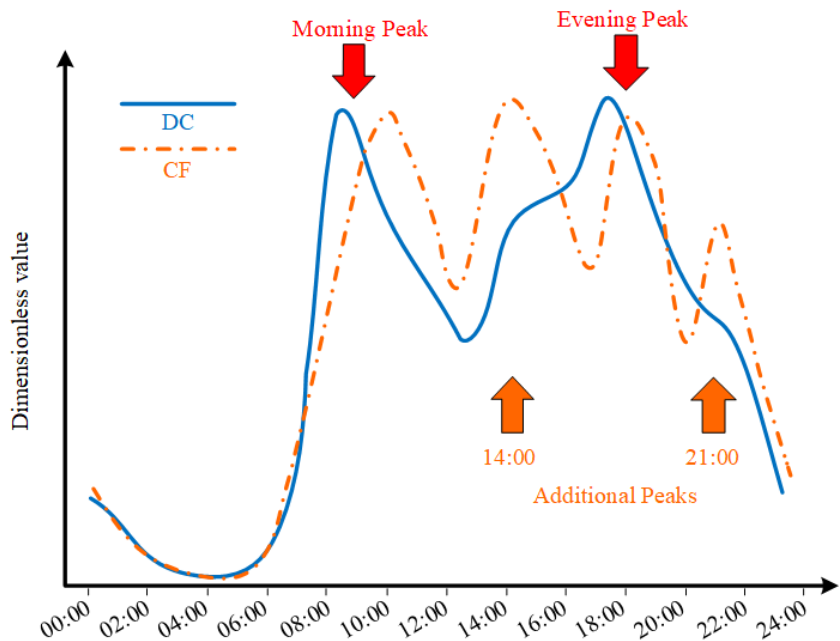

**Figure 7.** Temporal pattern comparison (daily sample)

To further strengthen the validity and applicability of this finding, an extra analysis was conducted based on GPS data from another city in China from the same period. The data contains 30 days of car-hailing vehicle traces of Xi'an city from October 1st to October 30th, 2016. Similar processing procedures were employed on the data and workday results are chosen to be presented in Figure 8.

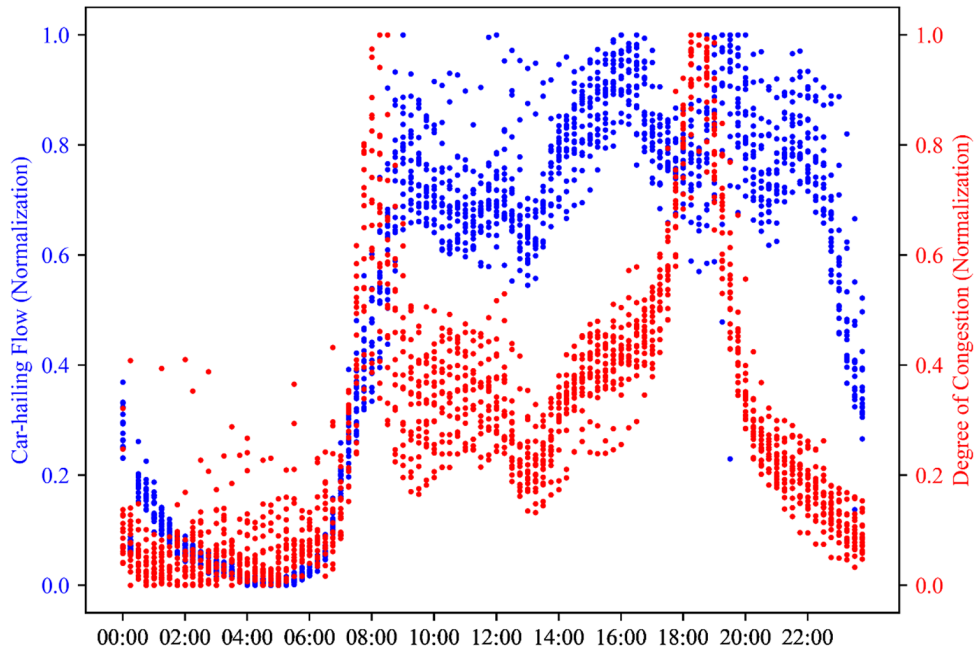

**Figure 8.** Workday CF and DC score values (extra city)

The extra experiment also demonstrates the differences between car-hailing vehicles and regular urban traffic, where 'multi-peaks'(blue points) can be observed in car-hailing vehicles and common 'morning-evening double peaks' (red points) can be observed in urban traffic. The mult-peaks may be shifted either front or back of time due to city-specific factors (such as geographical layout, traffic regulations, and cultural influences on travel behaviors); however, different city data confirms that car-hailing services in these two cities demonstrate similar patterns, primarily driven by the user demand for these services

*6.2 Spatial differences*

As for spatial scope, vehicles congregate at the same points, corresponding to the CBD or points of interest in Chengdu City. Nevertheless, differences still exist. As can be more clearly observed in Figure 6(a-d), car-hailing vehicles tend to travel along the ring roads to save time (the ring roads in Chengdu were designed as expressways). Moreover, Figure 7 reveals that the morning and evening peaks for car-hailing services occur slightly later than those for urban traffic. This time delay might manifest as postponed spatial aggregation in the spatial domain, a factor that must be considered when estimating urban traffic patterns using car-hailing trace data.

In conclusion, decisions concerning the timing and locations for utilizing car-hailing services impact car-hailing services, leading to a divergence from typical urban traffic patterns. Importantly, the spatial characteristics of car-hailing services are subject to long-term influences from drivers' driving habits and passenger demand. These spatial features remain consistent over extended periods and are not altered by longer input times, except when influenced by seasonal variations or during specific periods, such as the seven-day National Day holiday covered in this study.

*6.3 Potential usage values*

By its very nature, car-hailing vehicles are components of urban traffic flow, and there are intrinsic correlations between them. Understanding their differences in the temporal-spatial domain is crucial for developing effective management methods for transportation departments and operation approaches for car-hailing service providers. The proposed framework can simultaneously estimate the patterns of both car-hailing vehicles and urban traffic, enabling analysis of their temporal-spatial patterns for more flexible and cost-effective urban mobility. This analysis demonstrates the differences between car-hailing services and urban traffic flow in the temporal-spatial domain, offering two main potential usage values.

First, understanding these differences can provide essential supplemental information for transportation management staff and policymakers. In the big-data era, the transportation department considers car-hailing service data a major data source that is cost-effective and easy to obtain compared to traditional sensor data. Comprehending the different patterns of car-hailing services and urban traffic flow, and distinguishing the collected car-hailing data from actual urban traffic, will contribute to a good grasp of transportation network conditions. Moreover, it can help transportation management staff and policymakers formulate more precise policies and methods to alleviate traffic congestion when provided with car-hailing data.

Second, understanding the difference can also facilitate a comparison of different travel modes and induce travelers to choose car-hailing services(Wang et al., 2024), such as Mobility as a Service (MaaS) (Li et al., 2023; Wong et al., 2020) or Mobility-on-demand system (Huang et al., 2024). Urban traffic reflects private car conditions more, while car-hailing is another mode choice, a sharing economy (Barnes et al., 2020) that can reduce car ownership (Rayle et al., 2016). By leveraging the knowledge of the temporal-spatial difference between car-hailing vehicles and urban traffic, dispatching solutions or dynamic pricing policies for different periods and locations can be developed to offer better car-sharing mobility and attract more private car owners to use car-hailing services.

**7. Conclusions**

This paper presents an easy-to-handle approach for estimating and analyzing urban traffic patterns using car-hailing vehicle trace data. The proposed framework integrates essential functionalities, and a congestion score calculation procedure based on the INRIX index is employed to describe the status of roads. The effectiveness of the framework is evaluated through experiments conducted with real-world car-hailing record data, and temporal-spatial distribution analyses are carried out based on the estimated car-hailing flow matrix and urban traffic congestion score matrix.

The experimental results demonstrate the proposed framework's effectiveness for estimating and analyzing urban traffic patterns. The findings reveal a 'multi-peaks' temporal distribution of car-hailing vehicle flow and a bimodal temporal distribution of road congestion status. Moreover, a hierarchical structure caused by holiday traffic fluctuations is discovered and explored in detail. Spatial distribution analyses show that the hotspots of car-hailing vehicles and urban traffic exhibit spatial convergence and relative consistency in vehicle aggregations. The findings on temporal-spatial distribution patterns offer insights into investigating the relationships between car-hailing vehicles and urban traffic, and evaluating the effects of holidays on urban traffic patterns. This work also highlights the potential contributions of analyzing different patterns in car-hailing vehicles and urban traffic.

In this study, while the proposed framework effectively estimates urban traffic patterns using car-hailing data, it faces limitations regarding process and data quality handling. Particularly, the presence of incomplete or sparse estimated data may pose a risk to the reliability of the results. Developing tailored methods to enhance these tensors, possibly through data imputation techniques or machine learning models, is a promising direction for future research. Another promising research direction involves integrating the current framework with Hadoop to leverage its efficient capabilities in data processing and analysis.